\documentclass[%
 reprint,
superscriptaddress,
 amsmath,amssymb,
 aps,
pra,
]{revtex4-2}
\usepackage{babel}

\usepackage{graphicx}
\usepackage{dcolumn}
\usepackage{bm}
\usepackage{color}

\begin{document}
\preprint{APS/123-QED}

\title{Quantum boomerang effect of light}

\author{Xiangrui Hou}
\altaffiliation[]{Contributed equally}
\author{Zhaoxin Wu}
\altaffiliation[]{Contributed equally}
\author{Fangyu Wang}
\affiliation{School of Physics and Zhejiang Key Laboratory of Micro-nano Quantum Chips and Quantum Control, Zhejiang University, Hangzhou 310058, Zhejiang Province, China}
\author{Shiyao Zhu}
\affiliation{School of Physics and Zhejiang Key Laboratory of Micro-nano Quantum Chips and Quantum Control, Zhejiang University, Hangzhou 310058, Zhejiang Province, China}
\affiliation{Hefei National Laboratory, Hefei, 230088, China}
\author{Bo Yan}
\author{Zhaoju Yang}%
 \email{zhaojuyang@zju.edu.cn}
\affiliation{School of Physics and Zhejiang Key Laboratory of Micro-nano Quantum Chips and Quantum Control, Zhejiang University, Hangzhou 310058, Zhejiang Province, China}


\begin{abstract}
The quantum boomerang effect is a counterintuitive phenomenon where a wave packet, despite having an initial momentum, returns to its starting position in a disordered medium. However, up to now, the experimental exploration of this effect remains largely unexplored. Here, we report the experimental observation of the quantum boomerang effect of light. Our experiment is based on a one-dimensional disordered photonic lattice, which is composed of on-chip optical waveguides with engineered on-site random potential. We first characterize this optical disordered system by demonstrating the static Anderson localization of light beams. Next, through launching a kinetic light beam into the system, we observe that the light beam first moves away from its starting point, arrives at a maximum value, reverses its direction, and returns to its original position over time, confirming the observation of the quantum boomerang effect of light. Surprisingly, we find that optical loss, usually considered to be detrimental to optical experiments, can enhance the quantum boomerang effect by accelerating the light back to its original position. Our work provides new insights into the light-matter interactions in disordered medium and opens an avenue for future study of this phenomenon in nonlinear and many-photon contexts.
\end{abstract}

\maketitle
Controlling light has been at the forefront of modern optics for decades and is of fundamental interest to the photonic community. Advances in controlling light in photonics have led to the development of photonic crystals \cite{Yablonovitch1987, John1987} and metamaterials \cite{pendry2000negative, chen2010transformation}, which enable remarkable phenomena such as the inhibition of spontaneous emission within the bandgap and the steering of light with a negative refractive index \cite{fang2005sub, schurig2006metamaterial, cai2007optical}. More recently, the introduction of topology into photonics, inspired by topological insulators in condensed matter physics, has opened up new avenues for light control \cite{haldane2008possible, Wang2009, Fang2012, Khanikaev2013, rechtsman2013photonic, Hafezi2013, lu2014topological, ozawa2019topological, kim2020recent, fractal2022}. The realizations leveraging artificial gauge fields in gyromagnetic photonic crystals \cite{haldane2008possible, Wang2009}, helical waveguides \cite{rechtsman2013photonic}, silicon photonics \cite{Hafezi2013} et. al. offered the conventional photonic systems one-way topological edge states that are immune to imperfections and disorder. This field has opened new possibilities for robust light transport, unidirectional waveguiding, and even the realization of exotic topological phases that have no electronic counterpart, such as lasing \cite{Bahari2017, bandres2018topological, shao2020high, yang2020mode} and quantum photon sources \cite{mittal2018topological, dai2022topologically}. The further interplay between the topological phases and non-Hermitian physics \cite{ yao2018non, kawabata2019symmetry, bergholtz2021exceptional} has led to new frontiers in photonics \cite{guo2009observation,ruter2010observation, feng2013experimental, hodaei2014parity, feng2014single, zhen2015spawning, zeuner2015observation, weimann2017topologically, ni2018pt, el2018non, ozdemir2019parity, miri2019exceptional, zhao2019non, wang2021topological}. One paradigmatic example is the non-Hermitian skin effect \cite{yao2018edge, song2019non1, song2019non2, yokomizo2019non, yi2020non, zhang2020correspondence, yang2020non, xiao2020non, okuma2020topological, helbig2020generalized, Sun2024}, where extended bulk states condense at one edge, resulting in the funneling of light \cite{weidemann2020topological}. These advances underscore the potential for novel photonics that leverage the unique properties of topological and non-Hermitian systems.

\begin{figure*}[ht!]
\centering\includegraphics[width= 15cm]{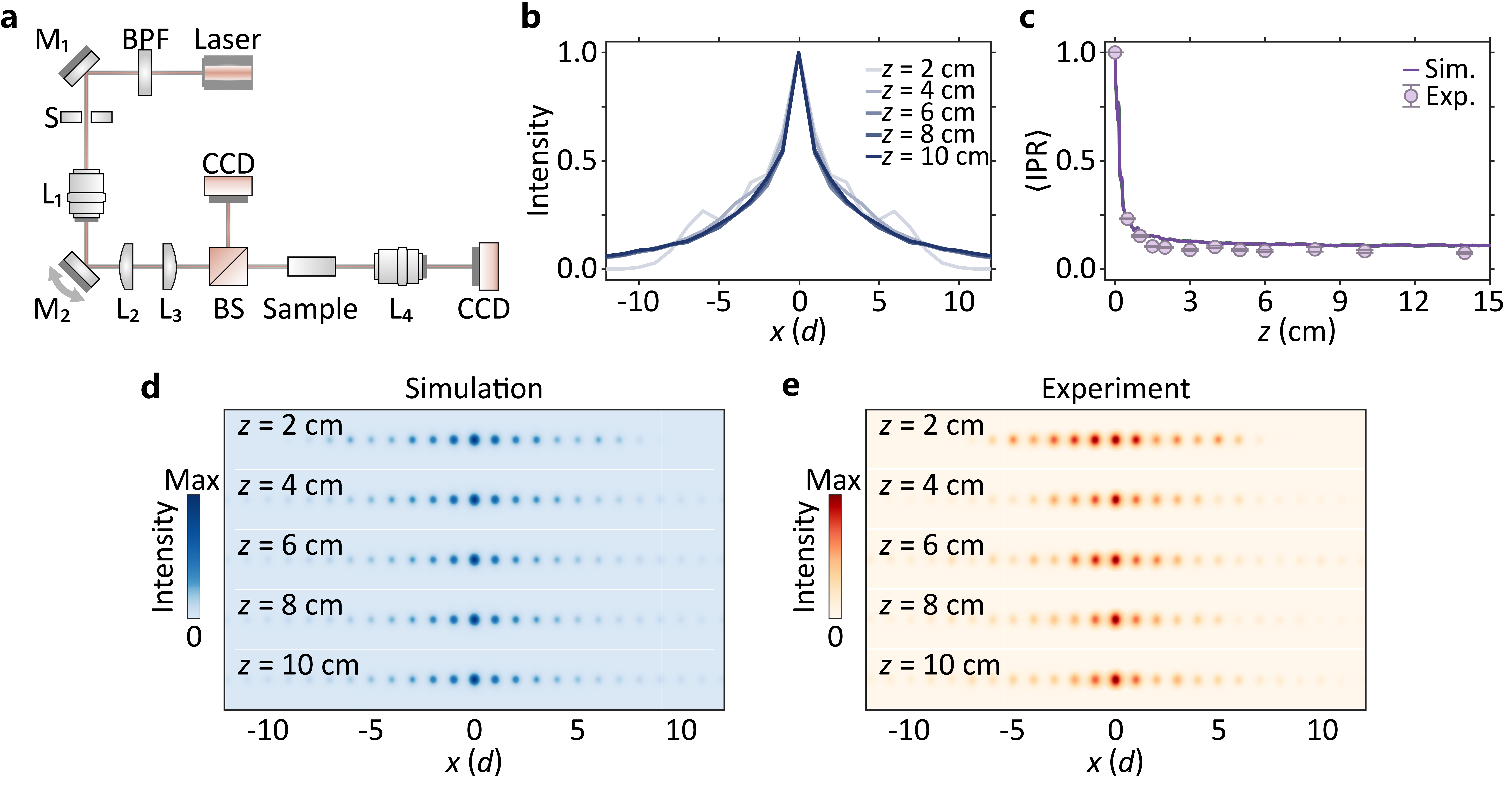}
\caption{Anderson localization of light in the 1D disordered optical lattice. \textbf{a}, The experimental characterization setup. (BPF: band-pass filter; M: mirror; S: slit; L: lens; BS: beam splitter; CCD: charge-coupled device.) \textbf{b}, The intensity profiles at several certain propagation lengths under single-site excitation. \textbf{c}, The IPR values at different propagation lengths. The solid line indicates the numerical results (under BPM simulation) and the dots indicate the experimental results. \textbf{d}, \textbf{e}, The intensity distribution of light at $z=2, 4, 6, 8$ and $10 \, \text{cm}$. Panel \textbf{d} (\textbf{e}) shows the numerical (experimental) results averaged over $10^3$ (21) realizations, respectively. The lattice constant is set to be $d=$ 15 \textmu m.}
\label{fig:1}
\end{figure*}

On the other hand, Anderson localization \cite{Anderson1958} is a phenomenon where waves become trapped in a disordered medium due to interference, preventing them from diffusing through the material. This concept was originally introduced by Anderson in 1958. However, nowadays Anderson localization has been experimentally observed in various systems beyond its original electronic context, including photonic lattices \cite{john1987strong, de1989transverse, Wiersma1997, Chabanov2000, Schwartz2007, lahini2008anderson, lahini2009observation, Levi2011, Levi2012, Sperling2013, Vardeny2013, Segev2013}, where light waves are confined by a disordered structure, and cold atom systems, where matter waves are localized \cite{Roati2008, Billy2008, Kondov2011, Jendrzejewski2012}. Building on this concept, the quantum boomerang effect - a counterintuitive dynamical feature beyond Anderson localization, was theoretically proposed \cite{prat2019quantum, janarek2020quantum, tessieri2021quantum, janarek2022quantum, noronha2022robust, Noronha2022, Janarek2023}, which disproves the common wisdom that, in a highly disordered lattice, waves can only be localized. In this phenomenon, a wave packet launched with an initial velocity in a disordered medium unexpectedly returns to its starting point over time, rather than dispersing or continuing along its trajectory. This intriguing phenomenon can be attributed to the combination of Anderson localization and parity-time symmetry of the system \cite{noronha2022robust, sajjad2022observation}. The former leads to a time-independent distribution of the wave packet at long times, and the latter leads to the wave packet's distribution being independent of the direction of momentum (Supplemental Material, Section I). While this effect has been extensively theoretically studied, its realization was only recently reported in the momentum space of ultracold atoms \cite{sajjad2022observation}, which limits the potential applications of this effect and the generalization to other systems such as optical and sonic waves. Given these developments, a compelling question naturally arises: can the quantum boomerang effect be realized with light? 

Here, we demonstrate the experimental observation of the quantum boomerang effect in photonics. Our starting point is the paraxial propagation of light in waveguides, where the propagation direction of axis \textit{z} plays the role of time \cite{rechtsman2013photonic}. Thus, this photonic system can be described by a Schrodinger equation for light mimicking the quantum behaviors of particles. First, we realize a one-dimensional (1D) disorder photonic lattice consisting of optical waveguides and observe the Anderson localization of light. Next, the interplay between the kinetic wave packet and disorder is investigated. In simulations, when the wave packet is initially launched with some momentum, we can see that the center of mass (COM) begins to move away from its initial position, arrives at a maximum value, reverses its direction, and returns to its original position over time, which manifests the quantum boomerang effect. Importantly, the engineered non-Hermiticity of gradient loss can enhance the boomerang effect by accelerating the COM back to its original position—a controlling mechanism not previously proposed in the literature. In our experiments, we directly observe the COM of the launched wave packet effectively returning to its starting point in both Hermitian and non-Hermitian disordered photonic lattices. A comparison between these two cases verifies the role of loss in modulating the COM's evolution. Our results not only deepen our understanding of light-matter interactions in disordered systems but also open up new avenues for controlling light in novel ways such as cloaking \cite{Pendry2006} and optical tweezers \cite{Grier2003}.

Our model is realized in an array of evanescently coupled waveguides \cite{rechtsman2013photonic, Sun2024}. The waveguides that can be directly written by femtosecond lasers are equally spaced with a lattice constant of $d$. The paraxial propagation of light in this system can be well described by the Schrodinger-type equation \cite{rechtsman2013photonic}:
\begin{equation}
    i\partial_{z}\Psi=- \frac{1}{2k_{0}}\nabla^{2}\Psi-\frac{k_{0}}{n_{0}}\Delta n\Psi,
\end{equation}
where $\Psi$ is the electric field amplitude, the Laplacian $\nabla^2$ acts on the transverse plane, $k_0=2\pi n_0/\lambda$ is the wavenumber in the background medium, and $\lambda$ is the wavelength of light. Hereafter, the ambient medium for our system is the fused silica with refractive index $n_0=1.45$, and $\Delta n(x,y,z)$ is the effective on-site potential, which is propagation-invariant and random in the transverse plane to construct a disordered photonic lattice. The paraxial propagation of light in this photonic system can be further governed by the tight-binding equation \cite{rechtsman2013photonic}:
\begin{equation}
    i\partial_{z}\psi_n=\sum_{\langle mn \rangle} c_0\psi_m + V_n \psi_n = H\psi_n,
\end{equation}
where $n\in[-N, N]$ is the lattice index, $\langle mn \rangle$ indicates the summation is taken over all the neighboring sites, $c_0$ is the coupling strength and $V_n$ represents on-site potential, which is set to be pseudo-random values \cite{griniasty1988localization} given by the form of $V_n^{(k)}=\delta \cos[\sqrt{5}\pi(n+(k-1)\times5)]^\alpha$, where $\delta$ represents the strength of disorder, $k$ represents the phase shift parameter in $k$th realization of disorder and $\alpha$ is set to be 3 hereafter \cite{griniasty1988localization}. Note that the detuning of the on-site potentials can be introduced by changing the writing velocity of the femtosecond laser (Supplemental Material, Section IX).

To investigate the quantum boomerang effect of light, we first show that our system of the disordered photonic lattice exhibits the Anderson localization when the wave packet with zero momentum is initially launched. In experiments, the 1D optical array of waveguides is fabricated by utilizing the femtosecond-laser writing method \cite{rechtsman2013photonic}. The optical characterization setup is shown in Fig. \ref{fig:1}a. By utilizing the beam-propagation method (BPM), the intensity profiles at the output facet with different propagation lengths are numerically simulated (See details in Supplemental Material, Section VIII) and the results are plotted in Fig. \ref{fig:1}b. In our simulations, the effective coupling strength is about $c_0=1.55 \, \text{cm}^{-1}$. The strength of the pseudo-random potentials is about $\delta=2.2 \, \text{cm}^{-1}$. The parameter of $\alpha$ is 3, which indicates that all the states in the system are localized \cite{griniasty1988localization}. We can see that the ensemble-averaged intensity distribution stabilizes at about $z=$4 cm and remains there over propagation length. The profile of the output beam becomes exponentially localized, which is direct evidence of the Anderson localization. Note that without disorder, the output profile exhibits ballistic transport and its width increases linearly with the propagation length (Supplemental Material, Section II).

\begin{figure}[ht!]
\centering\includegraphics[width= 7.5cm]{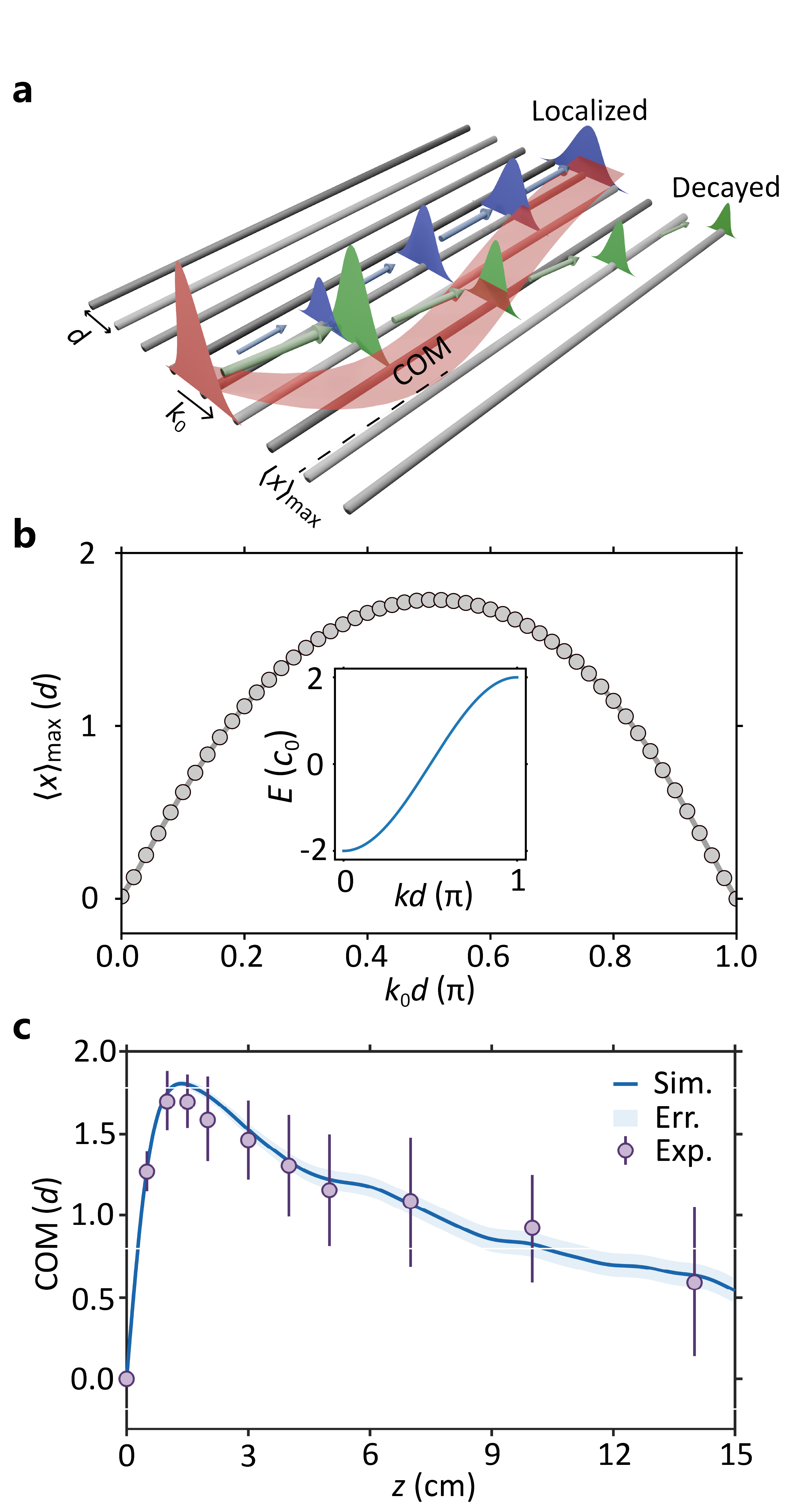}
\caption{Quantum boomerang effect of light in the 1D disordered optical lattice. \textbf{a}, Schematic of the quantum boomerang effect of light. The disordered optical lattice consists of equidistant waveguides with a lattice constant of \textit{d}. The varying grayscale represents the strength of the random potential. A wave packet with momentum \(k_0\) is launched into the lattice. The red color indicates the trajectory of COM. \textbf{b}, The relation between the momentum and the furthermost distance $\langle x \rangle_{\text{max}}$. The inset shows the corresponding energy band of a clean lattice. \textbf{c}, The COM at different propagation lengths. The solid curve and the shading indicate the averaged COM and the standard error (Err.) of the numerical results (tight-binding simulation). The dots indicate the experimental data. The parameters are set to be $c_0=1.55 \, \text{cm}^{-1}$, $\delta=2.2\, \text{cm}^{-1}$, $\sigma_0=3d$, and $N=100$. The numerical (experimental) results are averaged over $10^3$ (21) realizations. }
\label{fig:2}
\end{figure}

To experimentally confirm the Anderson localization in our photonic lattice, a laser beam with a wavelength of 633 nm is initially launched into the center waveguide of the 1D optical array. The measured intensity profiles at the output facet of the lattice are averaged over 21 realizations for the propagation length of $z=2, 4, 6, 8, $ and $ 10 $ cm, respectively. As we can see in Fig. \ref{fig:1}d $\&$ e, the output intensity profiles agree well with the simulation results, and the light beam becomes exponentially localized at about $z=4$ cm. We further characterize the localization behavior by using the inverse participation ratio (IPR) that is defined as $ \text{IPR}(z) \equiv \frac{\sum_n \left|{\psi_n}\right|^4}{(\sum_n \left|{\psi_n}\right|^2)^2} $. The numerical and experimental values of averaged IPR as a function of the propagation length of $z$ are depicted in Fig. \ref{fig:1}c. The IPR drops rapidly, stays stabilized at about $z=4$ cm, and maintains its stage throughout propagation, which indicates the static feature of the localization. Altogether, the numerical and experimental results reveal the ensemble-averaged intensity patterns undergo a transition from ballistic transport to localization when the disorder is introduced, which confirms that our disordered photonic lattice exhibits Anderson localization of light. 

Having presented the Anderson localization of light in our photonic lattice incorporating disorder, we next move forward to the interaction between the disordered lattice and the kinetic input wave packet with nonzero momentum. The kinetic beam can be expressed in a normalized Gaussian form of $\psi(x_n,z=0) \propto \exp\left(-\frac{x_n^2}{2\sigma_0^2}+ik_0 x_n\right)$, where $\sigma_0, k_0$ represent the width and momentum of the initial wave packet. Consequently, under the evolution of the photonic system, the COM of a launched wave packet as a function of propagation distance can be calculated by $ x(z)=\sum_{n}x_n|\psi(x_n,z)|^2$, and $\langle x \rangle$ indicates the ensemble-averaging of the COM over disorder realizations. The schematic of the quantum boomerang effect of light in our disordered lattice is shown in Fig. \ref{fig:2}a. A wave packet propagating in the disordered lattice can be divided into two parts: a localized part and a moving part. Initially, the moving part dominates, causing the COM to move away from its original position. As the wave evolves, however, the intensity of the moving part diminishes, and the localized part begins to take over. Consequently, the COM reaches a maximum value, denoted as $\langle x \rangle_{\text{max}}$, before eventually returning to its initial position over time. Even though the evolution of the COM of wave packets has been well explored in previous theoretical studies \cite{prat2019quantum, janarek2020quantum, tessieri2021quantum, janarek2022quantum, noronha2022robust, Noronha2022, Janarek2023}, the realization of the quantum boomerang effect in photonics is not that straightforward. In the following, we numerically simulate the model by sweeping experimentally accessible parameters and make sure that the quantum boomerang effect of light can be observed.

The question is: what is the underlying interplay between the momentum and the furthermost distance in the quantum boomerang effect? In the absence of disorder, the wave packet with a fixed momentum will acquire a group velocity that is governed by the dispersion relationship of the lattice. At the output facet at $z=15\,\text{cm}$, the relation between the momentum of $k_0$ and the maximum moving distance of $x_\text{max}$ is shown in Fig. S3 (Supplemental Material, Section III). We can see that the maximum traveling distance occurs at the momentum of $k_0 d=\frac{\pi}{2}$ corresponding to the dispersion with the maximum group velocity (see the inset in Fig. \ref{fig:2}b). Intuitively, we envision that this momentum induces the strongest boomerang effect. This is confirmed by numerical simulations, as shown in Fig. \ref{fig:2}b. The relationship between the $\langle x \rangle_\text{max}$ and the momentum reveals the wave packet with the initial momentum of $k_0 d=\frac{\pi}{2}$ travels the farthest before returning to its starting point. Therefore, we choose $k_0 d=\frac{\pi}{2}$ hereafter for the observability of the boomerang effect in experiments.

The disordered photonic lattices and the characterization setup are the same as that shown in Fig. \ref{fig:1}a. A broad Gaussian beam with about $\sigma_0=3\,d$ is generated after the laser beam with a wavelength of 635 nm passes through an optical slit. The momentum of the Gaussian beam is controlled by a rotated mirror and set to be $k_0 d=\frac{\pi}{2}$ (Supplemental Material, Section IX). The initial input beam is then launched into the sample, centering on the optical waveguide with the index of $n=0$. In order to trace out the trajectories of the launched wave packet, the intensity distributions at the output facet with the propagation length of 0, 0.5, 1.0, 1.5, 2, 3, 4, 5, 7, 10, and 14 cm are captured through a commercial CCD.

The measured pattern of the trajectory is shown in Fig. \ref{fig:2}c, with the solid line representing the numerical data and the dots indicating the experimental results. We can clearly see that the COM of the kinetic wave packet moves rapidly along the $x$ direction in the first z = 1 cm, reaches a maximum value of about $1.8\,d$, and then moves slowly to the starting position of $x=0$, which is the direct observation of the quantum boomerang effect of light. Note that due to the length limitation of the optical sample, the COM did not return completely to the starting position at z = 14cm.

\begin{figure}[ht!]
\centering\includegraphics[width= 7.5cm]{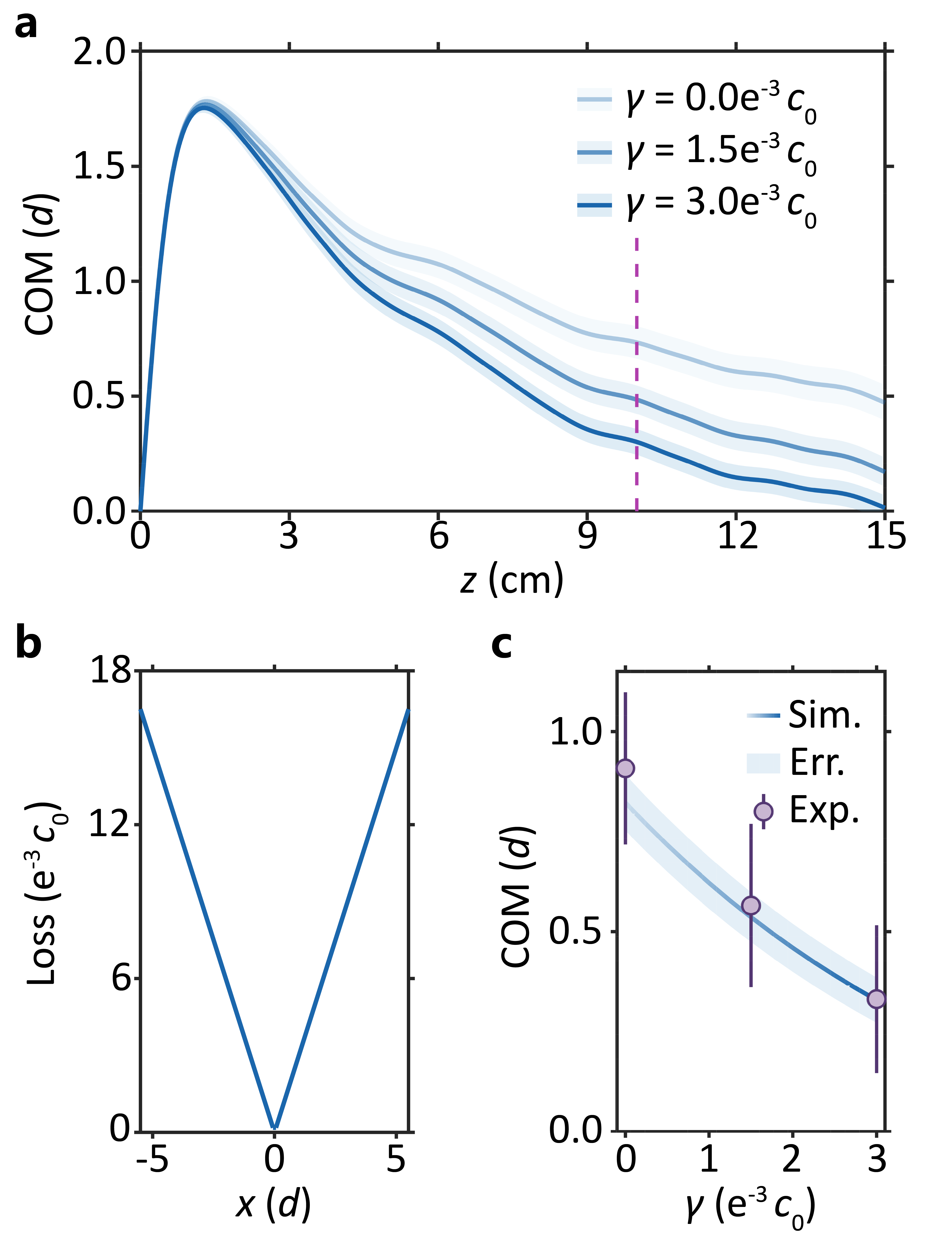}
\caption{Enhanced quantum boomerang effect by the symmetric gradient loss. \textbf{a}, The evolution of the COM with different strengths of gradient loss. \textbf{b}, The function of the gradient loss with $\gamma=3.0e^{-3}c_{0}$. \textbf{c}, The relationship between the COM and the strength of gradient loss at a fixed propagation length of $z=10$ cm (see purple dashed line in panel a). The depth of the blue color represents the strength of the gradient loss.} 
\label{fig:3}
\end{figure}

A natural question arises: can we accelerate the quantum boomerang effect?
Generally, loss is considered detrimental to optical experiments, particularly those involving photonic devices and quantum optics. However, surprisingly, we will show in the following that a specific design of loss configuration can facilitate the realization of the quantum boomerang effect by accelerating the return of the wave packet. The optical loss is configured to follow a symmetric gradient pattern, expressed as $-i\gamma |n| \psi_n^\dagger \psi_n$, where $n$, $\gamma$ represents the lattice index and the strength of gradient loss. The evolution of the COM of the launched wave packet is traced out by numerical simulations as shown in Fig. \ref{fig:3}a. The depth of the blue color indicates the gradient strength of $0$, $1.5e^{-3}c_0$, $3.0e^{-3}c_0$. We surprisingly find that a stronger gradient causes the COM to return to its starting position of $x=0$ more quickly, which facilitates the experimental realization of the quantum phenomena in our system. Note that the symmetric gradient loss alone cannot lead to the quantum boomerang effect in the photonic lattice without the disorder (Supplemental Material, Section II). In Fig. \ref{fig:3}b, we present the relationship between the strength of gradient loss and the COM at $z=10$ cm, which verifies that the introduction of gradient loss can accelerate the return of the wave packet. The experimental results agree well with the numerical ones. Note that we cannot apply infinitely strong strength of loss for acceleration, because in that case light decays too quickly to collect. 

\begin{figure}[ht!]
\centering\includegraphics[width= 7.5cm]{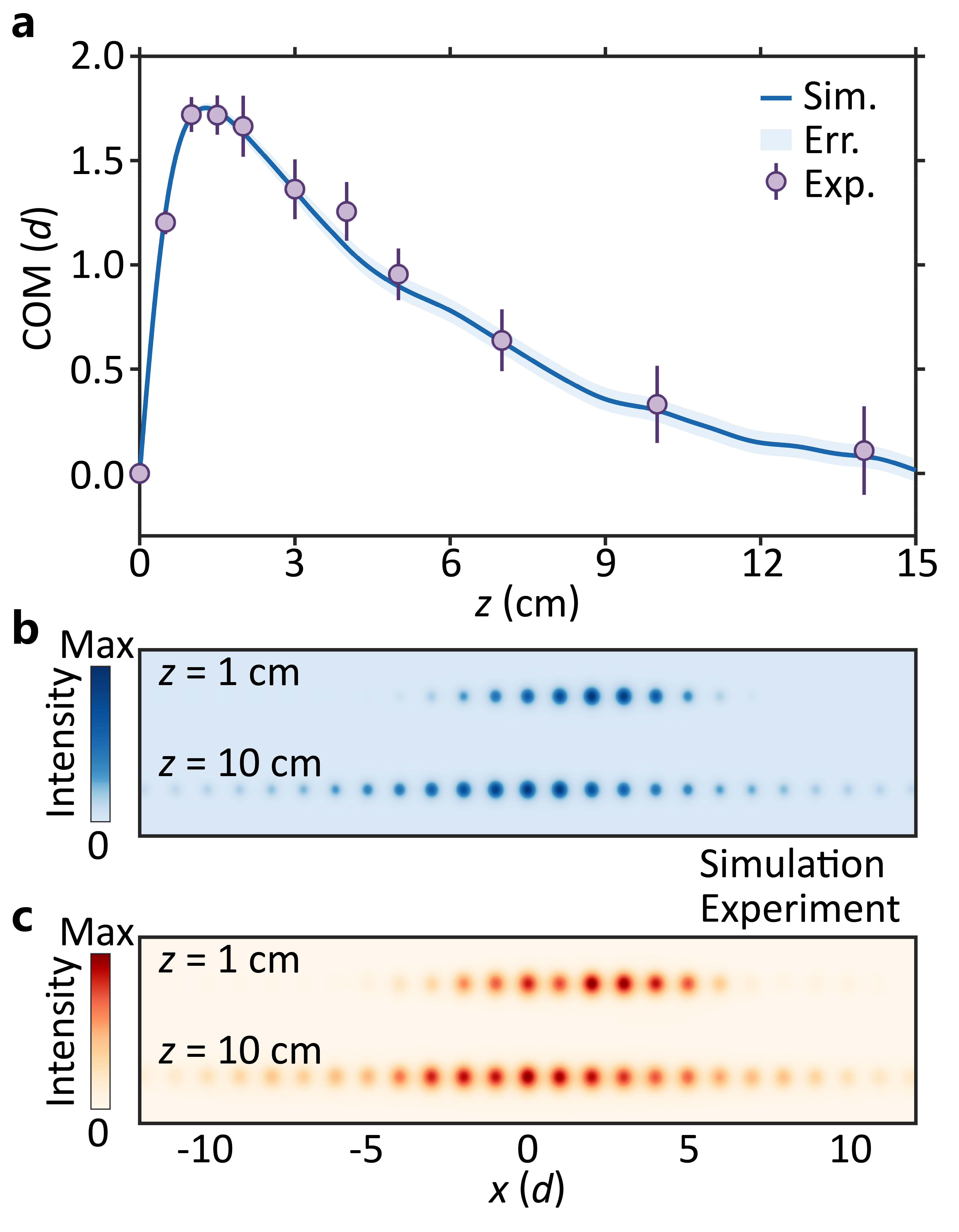}
\caption{Enhanced quantum boomerang effect of light. \textbf{a}, The COM at different propagation lengths. The solid curve and the shading indicate the averaged COM and the standard error of the numerical results from tight-binding simulations. The purple dots indicate the experimental data. \textbf{b, c}, The light intensity distribution at propagation length of $z=1$ cm and $z=10$ cm. The patterns are generated with the data from numerical simulations (BPM) and experiments. The numerical (experimental) results are averaged over $10^3$ (50) realizations. }
\label{fig:4}
\end{figure}

Following the same philosophy as previously, we now fabricate disordered photonic lattices with configured loss and perform the experiments. The gradient loss is introduced by setting breakpoints periodically in the optical waveguides \cite{Sun2024}. The gradient strength is set to be $\gamma=3.0e^{-3}c_0$. The measured pattern of the trajectory is shown in Fig. \ref{fig:4}a. The solid curve represents the numerical result from tight-binding simulations and the purple dots present the measured results. In Supplemental Material, Section IX, we also plot the results from the BPM simulations. They agree well with each other. By comparing this plot with that in Fig. \ref{fig:2}c, we can see that the evolution trajectory of the COM behaves similarly. However, the wave packet returns to the starting position significantly faster than before due to the presence of configured loss. In Fig. \ref{fig:4}b, we plot the intensity distributions at the propagation length of $z=1 \& 10$ cm from BPM simulations. The corresponding intensity distributions measured in experiments are shown in Fig. \ref{fig:4}c. In both simulations and experiments, the COM of the wave packet at the output facet of $z=1 \, \text{cm}$ is shifted right about $1.8\,d$. However, at the propagation length of $z=10$ cm, the COM completely returns to its original position of $x=0\,d$. As a whole, we have observed the quantum boomerang effect of light and found that the introduced configured loss can significantly accelerate the return of the kinetic wave packet.

In summary, we have demonstrated the quantum boomerang effect in a disordered photonic system. We first retrieved the Anderson localization of light in a femtosecond-laser written disordered optical array. Then, despite the initial momentum imparted to the light beam, the wave packet was observed to return to its initial position, manifesting the quantum boomerang effect, which is a key signature of the underlying interplay between quantum coherence and disorder. The introduced non-Hermiticity plays the role of dissipating the kinetic energy of the launched light and facilitating the observability of the quantum phenomenon \cite{Mo2022Imaginary, Junkai2024manybody}. Many intriguing questions arise, for example, the effect of nonlinearity \cite{Schwartz2007, jurgensen2023quantized} on the quantum boomerang effect of light, the behavior of entangled photons \cite{Lahini2010PRL,crespi_anderson_2013} in this system, or the interaction with quantum many-body physics. Thus, our results not only provide insights into light transport in disordered media but also open up new avenues for future research in photonics and quantum physics.\\

This research is supported by the National Key R\&D Program of China (Grants No. 2023YFA1406703 and No. 2022YFA1404203), National Natural Science Foundation of China (Grants No. 12174339), and Zhejiang Provincial Natural Science Foundation of China (Grant No. LR23A040003). Z. Y. thanks C. G. for the helpful discussions.

\nocite{*}

\bibliography{QBE_reference}

\end{document}